\documentclass[prl,twocolumn,preprintnumbers,amsmath,amssymb,superscriptaddress,showpacs]{revtex4}
\usepackage{graphicx}
\usepackage{latexsym}
\usepackage{amsmath}
\usepackage{graphics}
\usepackage{amssymb}
\usepackage{layout}
\usepackage{verbatim}
\usepackage{amsfonts,epsfig}
\newcommand{\ket}[1]{\left|#1\right>}
\newcommand{\bra}[1]{\left< #1 \right|}
\newcommand{\beq}{\begin{equation}}
\newcommand{\eeq}{\end{equation}}

\newcommand{\mean}[1]{\langle{#1}\rangle{}}

\begin{document}

\title{Dephasing of electron spin qubits due to their interaction with nuclei in quantum dots}
\author{{\L}ukasz Cywi{\'n}ski} \email{lcyw@ifpan.edu.pl}
\affiliation{Institute of Physics, Polish Academy of Sciences, al.~Lotnik{\'o}w 32/46, PL 02-668 Warszawa, Poland}
\date{\today }

\begin{abstract}
Coherence of spins of electrons confined in III-V quantum dots is strongly affected by their hyperfine interaction with the nuclei. In this paper an introduction to this subject is presented. Some theoretical approaches to the problem will be outlined. Most attention will be given to the Quasi-Static Bath Approximation, to the cluster expansion theories of dephasing due to the nuclear dynamics induced by the dipolar interactions (spectral diffusion), and to the effective Hamiltonian based theory of dephasing due to hyperfine-mediated interactions. The connections between the theoretical results and various experiments will be emphasized. 
\end{abstract}
\maketitle

\subsection{1. Introduction}

Coherently controlled electron spins embedded in a semiconductor environment have been a subject of intense research during the last ten years. The main motivation for the experimental and theoretical work was the prospect of their use as  qubits \cite{Cerletti_Nanotech05}, the basic building blocks of quantum computers. Recently, coherent control of a single spin was also shown to allow for very sensitive detection of magnetic fields with nanoscale resolution \cite{Maze_Nature08}.  

During the last few years a large progress was made in coherent control of spins in III-V materials such as GaAs \cite{Petta_Science05,Hanson_RMP07,Pioro_NP08,Koppens_PRL08,Barthel_PRL09,Foletti_NP09,Bluhm_NP11,Barthel_PRL10} and InGaAs \cite{Greilich_NP09,Press_NP10}. In these semiconductors all the nuclei are spinful, and the hyperfine (hf) interaction  with the nuclear spins is the main factor affecting the electron spin coherence time. Due to a large mismatch of Zeeman energies between the electron and the nuclei, the latter affect very little the longitudinal electron spin relaxation, which is thus due to phonon-induced transitions between the electronic states having a mixed spin character because of spin-orbit interaction, see e.g.~\cite{Khaetskii_PRB01,Golovach_PRL04,Hanson_RMP07}. Let us note that these interactions lead to spin relaxation (and spin dephasing \cite{Golovach_PRL04}) on timescales longer than miliseconds \cite{Hanson_RMP07}, which are much longer than the timescales considered here. 
The hf interaction with the nuclei is however the dominant source of the transverse spin decay, i.e.~the electron spin dephasing. Since the preservation of a specific phase relation (coherence) between the two states of the spin-$1/2$ is the necessary condition for the spin to be considered a qubit, the hf-induced decoherence needs to be understood.

The aim of this paper is to provide an introduction to the problem of hf-induced electron spin decoherence. In contrast to many other problems which involve decay of a population of a quantum state or  randomization of a relative phase between  two such states (e.g.~spontaneous recombination and dephasing of optical transitions), it is impossible use the Bloch equations \cite{Blum,Hu_arXiv01} to describe the dynamics of the electron spin interacting with the nuclei. In order to see this one should remember the conditions under which these equations are derived: the coupling between the system of interest (the electron spin here) and the bath should be weak (i.e.~the interaction should be small compared to the self-Hamiltonians of the spin and the bath), and the autocorrelation time of the bath should be much shorter than the timescale of the system's dynamics. Only then one obtains that the off-diagonal elements of the density matrix of the system decay according to $\rho_{ab}(t) \! \sim \! \exp(-t/T_{2})$. This is not the case for the electron spin interacting with the nuclei: the nuclear dynamics is very slow, and the coupling of the spin to the nuclei can be considered weak only in very high magnetic fields, $B\! \gg \! 1$ T, and only at these fields (and for a specially prepared state of the nuclei) the exponential decay of the transverse spin has been predicted \cite{Liu_NJP07,Coish_PRB08,Cywinski_PRL09,Cywinski_PRB09,Coish_PRB10}.

We will attempt to give an overview of theoretical approaches to this problem, with special focus on theories desribing the pure dephasing situation, in which the longitudinal relaxation of the spin is neglected, and only the dephasing (the decay of the spin component transverse to an applied magnetic field) is considered. These are  the theories of decoherence due to the dipolarly-induced dynamics of the nuclei (the so-called spectral diffusion) \cite{deSousa_PRB03,Witzel_PRB05,Witzel_PRB06,Yao_PRB06,Saikin_PRB07,Witzel_PRB08,deSousa_TAP09}, and the theories focusing on hf-induced dynamics described by an effective Hamiltonian (containing so-called hyperfine-mediated interactions) \cite{Yao_PRB06,Saikin_PRB07,Liu_NJP07,Cywinski_PRL09,Cywinski_PRB09}. The predictions of these theories for the spin echo decay in a GaAs QD have been recently confirmed experimentally \cite{Bluhm_NP11}.

The only analytical approach not manifestly of the pure dephasing kind which will be discussed here is the Quasi-Static Bath Approximation (QSBA) \cite{Dobrovitski_01,Merkulov_PRB02,Taylor_QIP06}, which is applicable at the timescale at which the nuclear dynamics is absent or can be treated in a very simplified manner (i.e.~by replacing all the nuclear spins by a collective classical vector). 
There are other approaches to the decoherence problem which \emph{do not} use the effective Hamiltonian, and with which the long-time decoherence was calculated \cite{Coish_PRB04,Deng_PRB06,Deng_PRB08,Coish_PRB10}. The relation between these theories and the effective Hamiltonian based approach is a subject of current research. However, since the spin echo signal has not been calculated using these methods, and since we want to emphasize here the connections between the theory and current experiments, we will not discuss these approaches further. The exact numerical studies of spin decoherence will also be mentioned rather briefly, and an interested reader is referred to a review \cite{Zhang_JPC07}.

Let us mention what else is \emph{not} covered in this paper. Most importantly, we do not explain how the spin initialization, manipulation, and readout are done experimentally. The III-V based spin qubits are controlled either electrically (by changing the voltages on the gates defining a QD in a two-dimensional electron gas), or optically (by applying short, coherent, and properly shaped pulses of light resonant with optical transitions in self-assembled QDs). The physics of each of the control schemes is a huge topic in itself, and the interested reader is referred to review papers (e.g.~\cite{Hanson_RMP07} for electrically controlled QDs, and \cite{Economou_springer10,Liu_AP10} for the optically controlled ones). 
A related omission is the lack of discussion of intricate physics of the singlet-triplet qubit realized in gated QDs (in this qubit the two states of interest are the singlet and unpolarized triplet states of two electrons residing in two coupled QDs). For simplicity, we focus on the case of a single spin interacting with a nuclear bath, but we note that all of the results on decoherence are, after suitable modifications, applicable to the singlet-triplet qubit (see e.g.~\cite{Koppens_PRL07} for the case of Rabi oscillations and \cite{Bluhm_NP11} for the case of spin echo decay). Finally, let us mention that recently spins of heavy holes confined in QDs have started to attract attention. The interesting features of hyperfine coupling between hole spins and nuclear spins are discussed in \cite{Fischer_PRB08,Fischer_PRL10}.

We focus  on III-V based QDs, but the theoretical approaches discussed here are also applicable (with  some modifications) to other systems in which a localized electronic spin interacts with the nuclei. Most importantly, the theory of decoherence due to the nuclear dynamics induced by the dipolar interactions (discussed in Section 5) was applied to the case of an electron bound to a phosphorus impurity in Si \cite{Witzel_PRB05,Witzel_PRB06,Witzel_AHF_PRB07,Saikin_PRB07,Witzel_PRL10} and it shows an excellent agreement with the experimental results on spin echo in this system \cite{Tyryshkin_PRB03,Tyryshkin_JPC06}. Let us note here that in the case of Si one can get rid of nuclear spins by means of isotopic purification (removing $^{29}$Si and leaving only the spinless $^{28}$Si). The coherence is then limited by interaction with remote electron spins of other donors \cite{Witzel_PRL10}, or with the spins of  the dangling bonds on Si/SiO$_{2}$ interface \cite{deSousa_PRB07}. Such purification is impossible in GaAs or InAs, since there are no spinless isotopes of Ga, As, and In.

The paper is organized in the following way. In Section 2 we introduce the Hamiltonian of the system (electron + nuclear bath), discuss the characterstic energy- and time-scales, and sketch the derivation of the effective Hamiltonian which is applicable at not-very-low magnetic fields (we believe it is applicable down to $\sim \! 10$ mT in GaAs dots, at least at timescales of current experimental interest). In Section 3 we briefly discuss the experiments in which the spin decoherence time is measured. Motivated by the slowness of the bath dynamics (mentioned in Section 2), in Section 4 we present theoretical results for electron spin dynamics obtained within the QSBA. The spin dephasing due to dipolarly-driven nuclear dynamics is then discussed in Section 5. Finally, the theory of spin decoherence due to hf-mediated interactions (which are the dominant sources of decoherence at small magnetic fields) is outlined in Section 6.

%%%%%%%%%%%%
%% BASIC FACTS
%%%%%%%%%%%%
\subsection{2. Basic facts about the nuclear bath}  \label{sec:bath}
The Hamiltonian of the electron spin interacting with the nuclear bath is given by 
\beq
\hat{H} = \hat{H}_{\text{Z}} + \hat{H}_{\text{dip}}  + \hat{H}_{\text{hf}}  \,\,  ,
\eeq
with the terms corresponding to Zeeman, dipolar, and hf interactions, respectively. The Zeeman term is written for the magnetic field $B$ in the $z$ direction as
\beq
 \hat{H}_{\text{Z}} = \Omega \hat{S}^{z} + \sum_{i} \omega_{\alpha[i]} \hat{J}^{z}_{i\alpha} \,\, ,  \label{eq:HZ}
\eeq
with the electron spin splitting $\Omega$ and nuclear spin splittings $\omega_{\alpha} \! = \! -\gamma_{\alpha}B$ for distinct nuclear species (i.e.~distinct elements or isotopes having gyromagnetic ratios $\gamma_{\alpha}$), and with $i$ being the site index (the nuclear species index $\alpha[i]$ is assigned randomly to possible sites while maintaining a given ratio of concentration of species). In the dipolar term we employ the secular approximation \cite{Abragam}, i.e.~we keep only the interactions which conserve the Zeeman energy:
\beq
\hat{H}_{\text{dip}}  =  \sum_{i\neq j} b_{ij} ( \hat{J}^{+}_{i}\hat{J}^{-}_{j} -2  \hat{J}^{z}_{i}\hat{J}^{z}_{j} ) \,\, , \label{eq:Hdip} 
\eeq
where the summation is over the nuclei $i$ and $j$ of the same species, and the couplings given by 
\beq
b_{ij} = -\frac{1}{4}\hbar\gamma_{i}\gamma_{j}\frac{1-3\cos^{2}\theta_{ij}}{r^{3}_{ij}} 
\eeq
where $r_{ij}$ is the distance between the two nuclei and $\theta_{ij}$ is the angle of $\boldsymbol{r}_{ij}$ relative to the $B$ field direction. For nearest neighbors we have $b_{ij} \! \sim\! 0.1 $ neV in GaAs (which corresponds to the timescale of $\hbar/b_{ij} \! \sim \! 10 $ ms). Together with the fact that the nuclear Zeeman energies of Ga and As nuclei at $B \! =\! 1 $ T are of the order of tens of neV (corresponding to less than a milikelvin), this leads to the conclusion that at typical experimental temperatures $T$ we have $(\hat{H}_{\text{Z}}+\hat{H}_{\text{dip}})/k_{B}T \! \ll \! 1$. The nuclear bath is thus described by a high-temperature density matrix $\hat{\rho}_{\text{B}} \sim 1$.

The hyperfine Hamiltonian is $\hat{H}_{\text{hf}} = \sum_{i} A_{i} \boldsymbol{\hat{S}}\cdot \boldsymbol{\hat{J}}_{i}$, which we will write as $\hat{H}_{\text{hf}} \! = \! \hat{V}_{\text{O}} \hat{S}^{z}+ \hat{V}_{\text{ff}}$, where we have defined the Overhauser operator
\beq
\hat{V}_{\text{O}}  =  \sum_{i}A_{i}\hat{J}^{z}_{i} \,\, ,  \label{eq:Ov}
\eeq
and the electron-nuclei flip-flop operator 
\beq
\hat{V}_{\text{ff}} =  \frac{1}{2} \sum_{i} A_{i} ( \hat{S}^{+}\hat{J}^{-}_{i} + \hat{S}^{-}\hat{J}^{+}_{i} )  \,\, .  \label{eq:Vsf}
\eeq
These two are playing very different roles in the process of electron decoherence at finite magnetic field. 
%\beq
%\hat{H}_{\text{hf}} = \sum_{i} A_{i} \boldsymbol{\hat{S}}\cdot \boldsymbol{\hat{J}}_{i} = \sum_{i} A_{i} [ \hat{S}^{z}\hat{J}^{z}_{i} + \frac{1}{2} ( \hat{S}^{+}\hat{J}^{-}_{i} + \hat{S}^{-}\hat{J}^{+}_{i} )  ] \,\, ,  \label{eq:Hhf}
%\eeq

In the above equations the hf couplings are  $A_{i} = \mathcal{A}_{\alpha[i]}f_{i}$, where $f_{i} \! \equiv  \! |\Psi(\boldsymbol{r}_{i}) |^{2}$, the squared modulus of the electron envelope function at the $i$-th nuclear site (with normalization to the primitive unit cell volume: $\int_{V} |\Psi(\mathbf{r})|^{2} d\mathbf{r}= \nu_{0}$). The hf energies for a nuclear species $\alpha$ are $\mathcal{A}_{\alpha} = \frac{2}{3}\mu_{0} \hbar^{2} \gamma_{S} \gamma_{J\alpha} |u_{\alpha}|^{2}$, where $\mu_{0}$ is the vacuum permeability, $\gamma_{S}$ and $\gamma_{J\alpha}$ are the electron and nuclear spin gyromagnetic factors, respectively, and $u_{\alpha}$ is the amplitude of the periodic part of the Bloch function at the position of the nucleus of $\alpha$ species (the normalization is $\int_{\nu_{0}} |u(\mathbf{r})|^{2}d\mathbf{r} \! = \! 1$). For Ga, As, and In atoms $\mathcal{A}_{\alpha} \! \approx 30-50$ $\mu$eV (see e.g.~\cite{Cywinski_PRB09} or \cite{Coish_pssb09} and references therein). The largest hf coupling is $A_{\text{max}} \! \sim \! \mathcal{A}/N$, with $N$ being the effective number of nuclei in the dot (the number of the nuclei which are appreciably coupled to the electron). We use a common definition of $N \! \equiv \! \sum_{i}f_{i} / \sum_{i} f^{2}_{i}$, which leads to a relation $\sum_{i\in\alpha} A^{2}_{i} \! =\! n_{\alpha}\mathcal{A}^{2}_{\alpha}/N$, where $n_{\alpha}$ is the number (per unit cell) of nuclei of $\alpha$ species. 
%(example for a Gaussian wavefunction ????)

It is crucial to notice that due to a large difference in electronic and nuclear magnetons, in a finite $B$ field we have $\Omega \approx 10^{3} \omega_{\alpha}$. Because of this Zeeman energy mismatch a flip-flop between the electron and a nucleus is practically prohibited. It can only occur when the magnetic field is so low that $\Omega$ becomes comparable to the dipolarly-broadened nuclear linewidth. This linewidth in III-V materials is of the order of a few kHz \cite{Shulman_PR58}, which corresponds to magnetic fields of $\sim \! 0.1$ mT. At much larger fields the direct electron-nuclear spin flip-flop is forbidden, and one can take into account the effect of $\hat{V}_{\text{ff}}$ term perturbatively. 

In the second order of perturbation theory we have processes in which the electron spin flip-flops with the $i$-th nuclear spin (leading to a virtual state with energy differing by $\approx \! |\Omega|$ from the energy of the intial state), and then flip-flops back with the $j$-th nuclear spin. This leads to appearance of an effective \emph{hyperfine-mediated} interaction between the nuclei, which can be viewed as an analogue of the well-known RKKY interaction, only derived using a localized electronic wavefunction instead of Bloch waves. Formally, this interaction is derived by performing an approximate canonical transformation. The effective Hamiltonian is $\tilde{H} \! =\! e^{-\mathcal{\hat{S}}} \hat{H} e^{\mathcal{\hat{S}}}$ with an unitary operator $e^{-\mathcal{\hat{S}}}$ chosen in such a way that $\hat{V}_{\text{ff}}$ is removed from the Hamiltonian (for more details see \cite{Shenvi_scaling_PRB05,Yao_PRB06,Coish_PRB08,Cywinski_PRB09}). It should be noted that the transformation of states, $\tilde{\ket{\phi}}\! = e^{-\mathcal{\hat{S}}} \ket{\phi}$, which should accompany the transformation of the Hamiltonian, is \emph{not} performed in the theories of decoherence which are employing the effective Hamiltonian. This approximation is the price which has to be paid for obtaining a theory with a convenient structure.

In the lowest order in $\hat{V}_{\text{ff}}$ we have $\mathcal{\hat{S}} \! \approx \! \hat{V}_{\text{ff}}\hat{S}^{z}/\Omega$, and the resulting effective Hamiltonian contains the terms
\begin{eqnarray}
\tilde{H}^{(2)} & =  & -\sum_{i} \frac{A^{2}_{i}}{4\Omega} \hat{J}^{z}_{i}  + \hat{S}^{z}\sum_{i} \frac{A^{2}_{i}}{2\Omega} \Big(\hat{J}_{i}^{2} - (\hat{J}^{z}_{i})^{2} \Big)  + \nonumber \\ 
& & + \hat{S}^{z}\sum_{i\neq j} \frac{A_{i} A_{j}}{2 \Omega} \hat{J}^{+}_{i}\hat{J}^{-}_{j}  \,\, . \label{eq:H2}  
\end{eqnarray}
In this Equation the last term is the most important: it is the hf-mediated interaction. As expected, its strength is decreasing with increasing magnetic field. The crucial feature of this interaction is that it is \emph{long-ranged}: any two nuclei among the $N$ spins significantly coupled to the electron are coupled to each other with comparable strength. The role played by this interaction in electron spin dephasing will be discussed in Section 6. Let us also mention, that recent experiments \cite{Reilly_PRL10} are suggesting that this interaction is also affecting the dynamics of nuclear polarization (nuclear spin diffusion)  at timescales much longer than the electron coherence time.

% SUMMARY OF GENERAL FACTS
%The timescales associated with dipolar or hf-mediated couplings between the nuclei are of the order miliseconds.
% CHECK!!!
%However, as will be discussed in Sections 5 and 6, the timescales of electron spin \emph{decoherence} due to the nuclear dynamics also involve $\mathcal{A}$ and $N$, and they are of the order of $0.1-100$ $\mu$s depending on the magnetic field and the dot size.

%%%%%%%%%%%
%% EXPERIMENTS
%%%%%%%%%%%
\subsection{3. Experiments in which the electron spin coherence is measured}  \label{sec:experiments}
The spin dephasing affects all the experiments involving coherent manipulation of the electron spin. We will outline now the basic experimental procedures used to gain information on decoherence of the electron spins. This will, however, be a theorist's description, in which we will assume that the electron spin is initialized (say in ``up'' direction), then well defined perfect rotations of this spin are driven by external stimuli at prescribed moments of time, and finally the measurement (along a given axis) is done. Performing each of these steps for spins in quantum dots was in fact a major achievement of the last 10 years, involving feats of experimental physics in the lab, and often requiring new theoretical ideas. Since we cannot give justice to these topics here, we refer the interested reader to review papers: about electrically controlled spins in gated QDs one can learn more in \cite{Hanson_RMP07} (for the most recent experimental advances see e.g.~\cite{Foletti_NP09,Barthel_PRL10}), while the optical control of spin rotations is described in \cite{Economou_springer10,Liu_AP10}.
% Below we will explain the idealized experiments, and summarize the results obtained in III-V QDs.

%%%%%%%%%%%
%% FID
%%%%%%%%%%%
\subsubsection{Free Induction Decay}
Conceptually the simplest experiment is the Free Induction Decay (FID), or simply the free evolution of the spin. We assume that the spin is initially oriented in, say, $\boldsymbol{x}$ direction (with $B$ field, when nonzero, directed along the $\boldsymbol{z}$ axis). The spin is allowed to freely evolve for time $t$.  At the final time $t$ a measurement, say the projection of the spin on the $\boldsymbol{x}$ axis, is made. If the spin was truly free (i.e.~not interacting with any bath), its evolution would simply be a precession, and the FID signal recorded as a function of the evolution time would be $\mean{\hat{S}_{x}(t)} \! = \! \frac{1}{2}\cos \Omega t$. 

Such a simple oscillation is not observed for electron spins interacting with the nuclear bath. The reason is the spread of the Overhauser fields $\hat{V}_{\text{O}}$ in the ensemble of nuclear states, i.e.~the inhomogeneous broadening.
The currently made FID experiments are either spatial \cite{Greilich_Science06} or temporal ensemble \cite{Koppens_Science05,Johnson_Nature05,Petta_Science05,Barthel_PRL09}  measurements. In the first case many spins (each interacting with a different nuclear bath) are measured simultaneously, while in the second case one measures repeatedly the same spin, but the data acquisition time is long enough for the nuclear bath to appreciably change its state. Assuming ergodic dynamics of the nuclei the two cases are equivalent, and they correspond to averaging of the electron spin evolution over the thermal ensemble of the nuclear states. This averaging will be done in Section 4. Here let us only say that it is enough to assume the nuclei static in order to obtain a very fast decay of the FID signal on the timescale of $T^{*}_{2} \! \sim \! \sqrt{N}/\mathcal{A}$, which is about $10$ ns in typical III-V dots.

The $T^{*}_{2}$ time is an ensemble quantity. In a quantum computer one will deal with single qubits, and the single-spin decoherence time, $T_{2}$, will be more relevant. Such a decay time would be observed if one could acquire enough signal from a single spin on a timescale shorter than the time on which the Overhauser field fluctuates appreciably (this timescale is much longer than microseconds in GaAs dots \cite{Reilly_PRL08,Barthel_PRL09}), or if one measured an ensemble of spins having the same Overhauser shift (as in \cite{Greilich_Science06}). In such cases one would be dealing with a so-called narrowed state of the nuclear ensemble. The free evolution experiment with such a nuclear ensemble is called the narrowed FID (NFID). 

The theory behind various ideas for achieving such a narrowed state is beyond the scope of this review (see e.g.~\cite{Coish_pssb09} and references therein). As for the experiments, the NFID measurement was done on an optically driven ensemble of InGaAs QDs \cite{Greilich_Science06}, in which the decay time of $T_{2} \! \approx \! 3$ $\mu$s was seen. Recently a substantial narrowing of the nuclear state was also obtained in electrically controlled GaAs dots \cite{Bluhm_PRL10}, leading to a tenfold increase (compared to $T^{*}_{2}$) of the decay time. A progress in single-shot measurement og GaAs spin qubits was also made \cite{Barthel_PRL09}. 
%ALSO \cite{Reilly_Science08} ? - Bluhm writes that there's another explanation for this experiment...

%%%%%%%%%%%
%% RABI
%%%%%%%%%%%
\subsubsection{Rabi Oscillations}
A two level system with level splitting $\Omega$, when exposed to a field of amplitude $R$ coupling its levels and oscillating with frequency $\nu$, i.e.~described by the Hamiltonian
\beq
\hat{H} = \Omega \hat{S}^{z} + R\hat{S}^{x}\cos \nu t\,\, ,  \label{eq:HRabi}
\eeq
will exhibit Rabi oscillations of the occupancy of its levels (see e.g.~\cite{Scully}). These oscillations have the frequency $\Omega_{R} \! = \! (R^{2} + \Delta^{2})^{1/2}$, where $\Delta \! =\! \Omega-\nu$ is the detuning, and their amplitude is $R^{2}/\Omega^{2}_{R}$. When the coupling between the system and its environment can be treated in Markovian approximation (i.e.~when we can use Bloch equations), these oscillations are damped by $\exp(-t/T_{2})$ factor. While this result is inapplicable to the case of the electron spin interacting with the nuclear bath, the damping of the oscillations is of course expected. The Rabi oscillations were observed in electrically driven singlet-triplet GaAs qubit \cite{Petta_Science05}, and in a single spin qubit \cite{Koppens_PRL07,Nowack_Science07,Pioro_NP08}.
In \cite{Koppens_PRL07} the oscillations were visible for times up to a microsecond. A model explaining their decay will be briefly discussed in Section 4.

%%%%%%%%%%%
%% SE
%%%%%%%%%%%
\subsubsection{Spin Echo}
The inhomogenous broadening, which obscures the interesting quantum dynamics of the single spin decoherence in the FID experiment, can be removed by a Spin Echo (SE) sequence \cite{Abragam}, in which the electron spin is rotated by angle $\pi$ around one of the in-plane axes at the midpoint of its evolution. Such a protocol can be denoted as $t/2-\pi-t/2$: two free evolution periods with a fast external pulse in the middle and readout at the final time $t$. It is easy to see that such a procedure will remove the static spread of the precession frequencies, since the evolution of every spin (assumed free, only having a random precession frequency) before the $\pi$ pulse will be undone after the pulse. This refocusing of the spins does not work perfectly when the bath is dynamic, so the amplitude of the SE signal is still decaying in time. In fact, if the Bloch equations were applicable, the SE would decay as $\exp(-t/T_{2})$, with $T_{2}$ being the single spin dephasing time (the same would then be seen in NFID). In the more complicated case considered here, the decay of SE is non-exponential, and the characteristic time, $T_{\text{SE}}$, needs not be related to the decay time of NFID.

SE experiments were done in gated double GaAs dots, both for the singlet-triplet qubit \cite{Petta_Science05,Bluhm_NP11,Barthel_PRL10} and for the single spin in one of the dots \cite{Koppens_PRL08}. In the earlier experiments \cite{Petta_Science05,Koppens_PRL08} the magnetic field was $B \! \leq \! 0.1$ T, and the decay time was $T_{\text{SE}} \! \leq \! 1$ $\mu$s. This decay timescale was later explained by the theory \cite{Cywinski_PRL09,Cywinski_PRB09}, which also predicted a characteristic oscillatory behavior of the SE signal at $B \! > \! 0.1$ T. This prediciton has been recently confirmed \cite{Bluhm_NP11}, as discussed in more detail in Section 6.

Let us also mention that recently the SE experiment was performed with optically induced $\pi$ rotations on an ensemble of self-assembled quantum dots \cite{Greilich_NP09}, and on a single electron spin bound to a donor \cite{Clark_PRL09}, or confined in  an InAs dot \cite{Press_NP10}.

%%%%%%%%%%%
%% DD
%%%%%%%%%%%
\subsubsection{Dynamical Decoupling}
Application of a single $\pi$ pulse during the qubit evolution removes the effects of inhomogenous broadening, and furthermore it affects the dynamics of the whole system (qubit$+$nuclei). The application of multiple $\pi$ pulses is well known in NMR \cite{Abragam}, where these pulses were shown to further suppress spin dephasing and relaxation. In the context of quantum computation these ideas were furhter developed under the name of \emph{dynamical decoupling} (DD), see e.g.~\cite{Viola_JMO04}. For example, the Carr-Purcell-Meiboom-Gill (CPMG) sequence, which can be written as $\tau/2-\pi-\tau-\pi-...-\pi-\tau/2$ (with $n$ pulses and the total sequence time $t\! = \! n\tau$, equivalent to SE for $n\! =\! 1$), was predicted theoretically to extend the coherence time of an electron spin in GaAs at high $B$ fields \cite{Witzel_PRL07}, and this prediction has been verified experimentally \cite{Bluhm_NP11}. Other DD sequences were also considered theoretically for the case of spin decoherence due to hf interactions \cite{Yao_PRL07,Witzel_CDD_PRB07,Lee_PRL08,Zhang_Dobrovitski_PRB07,Zhang_Viola_PRB08,Cywinski_PRB09}. Experimentally, in GaAs singlet-triplet qubits both CPMG \cite{Bluhm_NP11,Barthel_PRL10} and other sequences with unequal spacing of pulses were employed  \cite{Barthel_PRL10}.

%%%%%%%%%%%%%
%%% QSA
%%%%%%%%%%%%
\subsection{4. Quasi-static bath approximation}  \label{sec:QSA}
As we mentioned before, the spin decay seen in the FID experiment can be calculated assuming completely static nuclei. This is equivalent to treating the $\sum_{i}A_{i}\boldsymbol{\hat{J}}_{i}$ operator as a classical field $\boldsymbol{B}_{N}$. We neglect also the nuclear Zeeman energy and the dipolar interactions, and use the QSBA Hamiltonian $\hat{H}_{\text{QSBA}} \! =\! \Omega\hat{S}^{z} + \boldsymbol{B}_{N}\cdot\boldsymbol{\hat{S}}$.
For the unpolarized nuclear ensemble, in the limit of large number of nuclei $N$, one can use the Central Limit Theorem and derive the distribution of the $\boldsymbol{B}_{N}$ fields (see e.g.~\cite{Dobrovitski_01,Merkulov_PRB02}):
\beq
P(\boldsymbol{B}_{N}) = \frac{1}{(2\pi)^{3/2}\sigma^{3}} \exp\left( -\frac{B^{2}_{N}}{2\sigma^{2}} \right ) \,\, ,
\eeq
with
\beq 
\sigma^{2} \! = \! \frac{1}{3}\sum_{\alpha}J_{\alpha}(J_{\alpha}+1)\sum_{i\in\alpha}A^{2}_{i} = \frac{1}{3}\sum_{\alpha}J_{\alpha}(J_{\alpha}+1)n_{\alpha}\frac{\mathcal{A}^{2}_{\alpha}}{N}  \,\, ,
\eeq
where $J_{\alpha}$ is the length of the nuclear spin of species $\alpha$ (e.g.~$J\! = \! 3/2$ for all the nuclei in GaAs, and $J\! = \! 9/2$ for both isotopes of In).

The calculation of the FID signal is especially easy at high $B$ fields, when we can disregard the influence of transverse components of $\boldsymbol{B}_{N}$ and approximate the Hamiltonian by $(\Omega + B_{z})\hat{S}^{z}$. Assuming that the spin is initialized along the $\boldsymbol{x}$ direction we have
\begin{eqnarray}
\mean{\hat{S}_{x}(t)} & = & \frac{1}{2}\int P(\boldsymbol{B}_{N}) \cos\left((\Omega+B^{z}_{N})t\right ) \text{d}\boldsymbol{B}_{N} \,\, , \nonumber \\
& = & \frac{1}{2} \cos (\Omega t) e^{-\left(t/T^{*}_{2}\right)^2 } \,\, ,  \label{eq:FID}
\end{eqnarray}
where $T^{*}_{2} \! = \! \sqrt{2}/\sigma \! \sim \! \sqrt{N}/\mathcal{A}$ (e.g.~for a bath with a single nuclear species and $J\! = \! 1/2$ we have $T^{*}_{2} \! = \! \sqrt{8N}/\mathcal{A}$). For typical III-V dots with $N\! \approx \! 10^{6}$ nuclei we have $T^{*}_{2} \! \approx \! 10$ ns. The results for FID decay obtained within QSBA were confirmed by experiments \cite{Dutt_PRL05,Koppens_Science05,Johnson_Nature05,Braun_PRL05}.

\begin{figure}[t]
\centering
\includegraphics[width=0.99\linewidth]{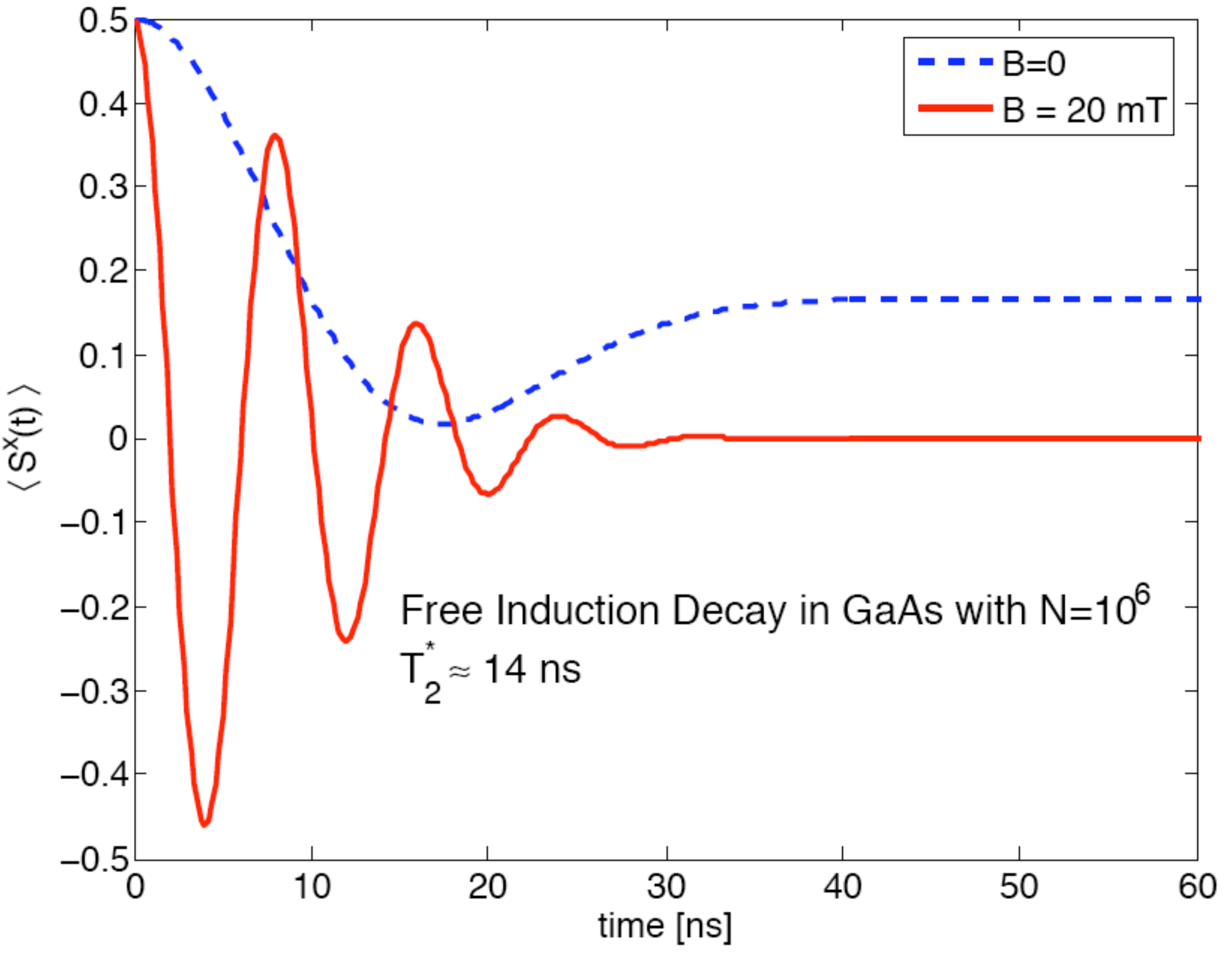}
  \caption{Free Induction Decay (FID) of the $S^{x}$ component of the electron spin in a GaAs dot with $N\! = \! 10^{6}$ nuclei. The calculations are for $B\! =\! 0$ and at $B$ much larger than the rms of the Overhauser field, specifically $B\! =\! 20$ mT. These calculations are done using Eqs.~(\ref{eq:FIDB0}) and (\ref{eq:FID}), respectively. At longer times the $B\! = \! 0$ signal is predicted to decay as $1/\ln t$ \cite{Erlingsson_PRB04,Al_Hassanieh_PRL06,Zhang_PRB06,Chen_PRB07}. The FID decay signals such as shown in this Figure have been measured in many experiments \cite{Koppens_Science05,Johnson_Nature05,Braun_PRL05,Petta_Science05}. } \label{fig:FID}
\end{figure}

The calculation is slightly less trivial in the case of $B\! =\! 0$, where has to average the expression for the time-dependence of the spin in a classical static field 
\beq
\boldsymbol{S}(t) \! =\! (\boldsymbol{S}_{0}\cdot\boldsymbol{n})\boldsymbol{n} + ( \boldsymbol{S}_{0}-(\boldsymbol{S}_{0}\cdot\boldsymbol{n})\boldsymbol{n})\cos B_{N}t + \boldsymbol{S}_{0}\times\boldsymbol{n} \sin B_{N} t \,\, ,  \label{eq:Sdyn}
\eeq
with the initial spin $\boldsymbol{S}_{0}$ and the direction of the effective field $\boldsymbol{n} \! \equiv \! \boldsymbol{B}_{N}/B_{N}$, obtaining \cite{Dobrovitski_01,Merkulov_PRB02}
\beq
\mean{\hat{S}_{x}(t)} = \frac{1}{6} \left( 1-2 \left(2\left(\frac{t}{T^{*}_{2}}\right )^2 -1\right) e^{-\left(t/T^{*}_{2}\right)^2 } \right) \,\, .  \label{eq:FIDB0}
\eeq
A plot of this result for GaAs with $N=10^{6}$ nuclei is shown in Fig.~\ref{fig:FID}.
Such a FID signal, which saturates at $\mean{\hat{S}_{x}(t)} \! = \! 1/6$ for $t\! > \! T^{*}_{2}$ was observed in \cite{Braun_PRL05}. The maintaining of this saturation at long times is an artifact of QSBA, which has to break down  eventually. Various theoretical approaches were used to predict the long-time $1/\ln t$ decay of the FID signal in this case \cite{Erlingsson_PRB04,Al_Hassanieh_PRL06,Zhang_PRB06,Chen_PRB07}. 

%%%%%%%%%%%%%%
%% RABI WITH QSA
%%%%%%%%%%%%%%
While QSBA is surely valid on a timescale of $T^{*}_{2}$, it is not known what is the precise limit on the timescale on which it is quantitatively valid (for a given experiment). One experiment, which sheds some light on this question, is the measurement of the decay of Rabi oscillations \cite{Koppens_PRL07}. 

In the rotating frame (in which the wavefunctions are transformed by the unitary operator $\exp(i\nu\hat{S}^{z}t)$), and within the Rotating Wave Approximation (in which the strongly oscillating terms are dropped from the Hamiltonian), we have the effective QSBA Hamiltonian
\beq
\hat{H}_{R,QSBA} = (\Delta + B^{z}_{N})\hat{S}^{z} + R\hat{S}^{x} \,\, . \label{eq:HRQSA}
\eeq
Assume that the spin is initialized in the ``up'' state at $t\! = \!0$, and that the detuning $\Delta \! =\! 0$. Furthermore, for simplicity let us assume $R \! \gg \! B^{z}_{N}$ (this does not affect the qualitative features of the result). The probability of finding the spin in the ``down'' state is given by \cite{Dobrovitski_01,Taylor_QIP06,Koppens_PRL07,Hanson_Science08}
\begin{eqnarray}
\mean{P_{\downarrow}(t)} & = & \int_{-\infty}^{\infty} P(B^{N}_{z}) \sin\left( Rt + \frac{(B^{z}_{N})^2}{2R} \right) \text{d}B^{z}_{N}\\
& = & \frac{ \sin\left( Rt + \frac{1}{2}\arctan \frac{\sigma^{2}t}{R} \right)}{\left(1+\frac{\sigma^4}{R^2}t^2 \right )^{1/4}} \,\, .
\end{eqnarray}
The asymptotic $\sim 1/\sqrt{t}$ decay predicted by the above formula was seen in GaAs spin qubit for times up to a microsecond \cite{Koppens_PRL07}, suggesting that QSBA is still valid at this timescale. The long-time $\pi/4$ phase shift of the oscillation was also visible in the data from \cite{Koppens_PRL07}. The power-law decay of the Rabi oscillation was also seen in a Nitrogen Vacancy center spin qubit in diamond \cite{Hanson_Science08}.

%%%%%%%%%%%%%%
%% NARROWED FID WITH QSA
%%%%%%%%%%%%%%
A calculation very similar to the one shown above can be performed for the case of the narrowed FID decay. Assuming for simplicity the narrowing condition $B^{z}_{N} \! =\! 0$, and assuming $B_{N} \! \ll \! \Omega$ (which corresponds to $\Omega \! \gg \! \sigma \! \approx \! \mathcal{A}/\sqrt{N}$), by averaging Eq.~(\ref{eq:Sdyn}) over the Gaussian distribution of $B^{x}_{N}$ and $B^{y}_{N}$ we obtain an expression very similar to the previous one:
\beq
\mean{S^{x}(t)} = \frac{\cos \Omega t - \eta t \sin\Omega t}{2(1+\eta^{2}t^{2})} = \frac{ \cos\left(\Omega t + \arctan \eta t\right)}{2\sqrt{1+\eta^{2}t^{2}}} \,\, , \label{eq:SxFID_QSA}
\eeq
where we have defined $\eta \! \equiv \! \sigma^{2}/\Omega$. Note that the characteristic decoherence time $T_{\text{NFID}}$, defined as the time in which the signal drops by half, is $T_{\text{NFID}} \! \approx \! N\Omega/\mathcal{A}^2$, and it increases with increasing $B$ field and dot size.

\begin{figure}[t]
\centering
\includegraphics[width=0.99\linewidth]{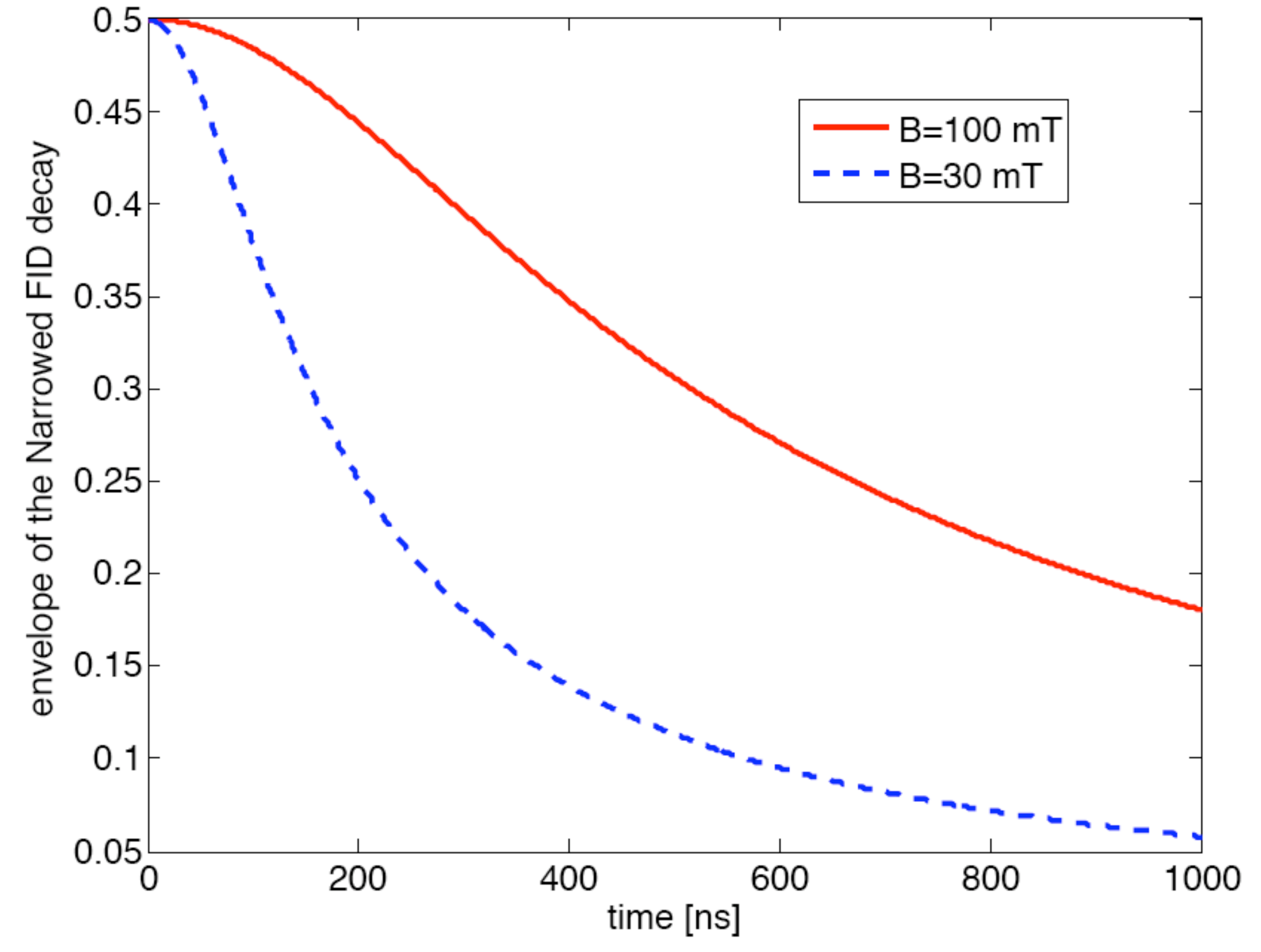}
  \caption{The envelope of the Free Induction Decay for a narrowed nuclear state (NFID) in a GaAs dot with $N\! = \! 10^{6}$ nuclei. The calculation is done using the QSBA (Eq.~(\ref{eq:SxFID_QSA})). The results are shown for $B\! =\! 30$ mT and $B\! = \! 100$ mT. At higher $B$ fields the decay becomes slower. At fields at which the decay timescale becomes longer than $N/\mathcal{A} \! \approx \! 10$ $\mu$s, the QSBA becomes inapplicable, and the theories (both the effective Hamiltonian based theory discussed in Section 6 and the Generalized Master Equation approach from \cite{Coish_PRB08,Coish_PRB10}) predict a mostly exponential decay.} \label{fig:NFID}
\end{figure}

In contrast to the Rabi oscillations discussed above, the envelope of the NFID signal decays asymptotically as $1/\eta t$, see Fig.~\ref{fig:NFID}. However, in the frame rotating with the $\Omega$ frequency, the decay is $\mean{S^{x}(t)} \sim 1/\eta^{2}t^{2}$ due to the long-time phase shift of $\pi/2$, which corresponds to the spin asymptotically rotating from the $\boldsymbol{x}$ axis to the $\boldsymbol{y}$ axis. This is an example of one of subtle effects associated with the bath-induced frequency shifts of the spin precession (see e.g.~Ref.~\cite{Coish_PRB10} for discussion of other such effects in a different theoretical framework).

In Section 6 we will show that Eq.~(\ref{eq:SxFID_QSA}) can be derived as a short-time ($t \! \ll \! N/\mathcal{A}$) limit of a theory which includes the nuclear dynamics. Since $N/\mathcal{A}$ is of the order of $10$ $\mu$s in GaAs dots, this suggests that the QSBA is applicable at such defined short times. This agrees with the observation of a good agreement between the QSBA and the Rabi experiment for times up to a microsecond \cite{Koppens_PRL07}. 

%%%%%%%%%%%%%%%
%%% DIPOLAR
%%%%%%%%%%%%%%
\subsection{5. High magnetic field limit - dephasing due to dipolar interactions}  \label{sec:dipolar}
In large enough $B$ fields one can completely neglect the flip-flop terms between the electron and the nuclei. The quantitative answer to the question how large these $B$ fields need to be will be provided in Section 6. Now let us focus on the consequences of such an approximation. With $\hat{V}_{\text{ff}}$ term neglected, we have the following Hamiltonian:
\beq
\hat{H} \approx \hat{H}_{\text{Z}} + \hat{H}_{\text{dip}} + \hat{V}_{\text{O}}\hat{S}^{z} \,\, .
\eeq
The $\hat{H}_{\text{dip}} $ is the only non-trivial term leading to nuclear dynamics. The crucial feature of this  Hamiltonian is that it is of the \emph{pure dephasing} form, i.e.~the only electron spin operator present in it is $\hat{S}^{z}$. The obvious consequence is the fact that $\hat{S}^{z}$ is then a constant of motion. A slightly less obvious fact is the structural simplification of the theory of decoherence which takes place in this case.  

The physical picture of transverse spin dephasing is the following.
The dipolar interaction leads to slow fluctuations of the Overhauser field: the nearby spins are flip-flopping because of the $\hat{J}^{+}_{i}\hat{J}^{-}_{j}$ term in Eq.~(\ref{eq:Hdip}), and due to the spatial inhomogeneity of the hf couplings (i.e.~$A_{i}\! \neq \! A_{j}$ for the two spins considered) this corresponds to a change of the expectation value of $\hat{V}_{\text{O}}$ by $A_{ij} \! =\! A_{i}-A_{j}$. This leads to fluctuations of the Overhauser field, and thus the fluctuations of the electron spin precession frequency.
It should be noted that the hf interaction plays a passive role in this process: the nuclear fluctuations are due to the dipolar interaction, and the Overhauser operator simply transmits these fluctuations to the electron spin.

\begin{figure}[t]
\centering
\includegraphics[width=0.99\linewidth]{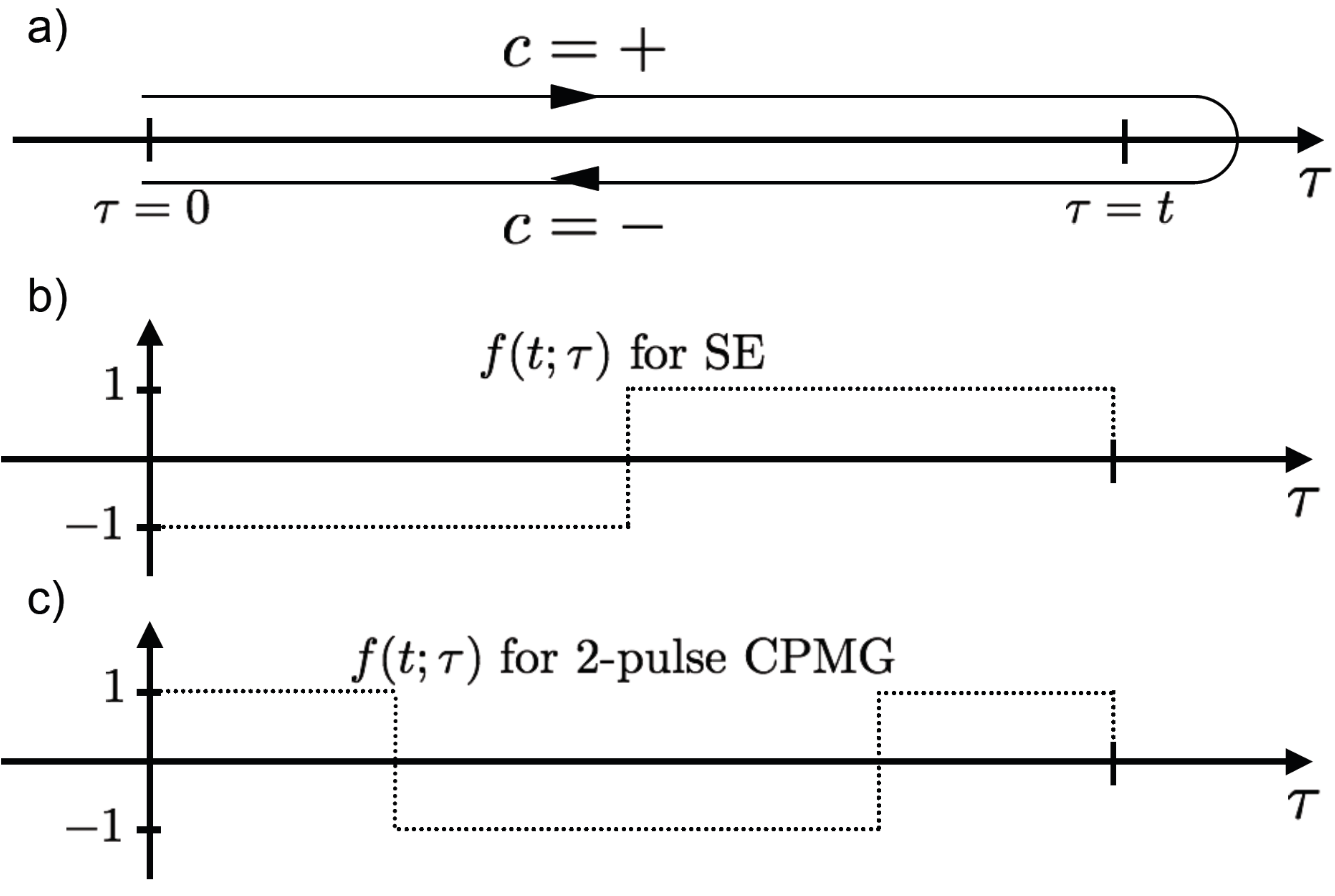}
  \caption{ a) The closed loop contour along which the operators in Eq.~(\ref{eq:WC}) are ordered. b)
 The plot of the time-domain filter function $f(t;\tau)$ for the Spin Echo sequence. c) The same for 2-pulse CPMG sequence. The Figure is adapted from Ref.~\cite{Cywinski_PRB09}.} \label{fig:contour}
\end{figure}

This process is known as \emph{spectral diffusion} (i.e.~the diffusion of the precession frequency of the spin). For many years it was treated using phenomenological classical methods, i.e.~by replacing  the operator $\hat{V}_{\text{O}}$ by a classical stochastic process $\xi_{\text{O}}(t)$ (see \cite{deSousa_PRB03} and references therein). The average transverse spin is then given by
\beq
\mean{\hat{S}^{x}(t)} = \text{Re} \left( e^{-i\Omega t} \mean{e^{-i\int_{0}^{t} \xi_{\text{O}}(t')f(t;t') \text{d}t'}   }_{\xi}  \right )  \,\, , \label{eq:classical}
\eeq
where $\mean{...}_{\xi}$ denotes the average over the realizations of the stochastic process $\xi_{\text{O}}(t)$, and $f(t;t')$ is a \emph{filter function} \cite{deSousa_TAP09,Cywinski_PRB08} which parametrizes the sequence of $\pi$ pulses applied to the qubit (for FID $f(t;t')\!= \! 1$, and the filters for SE and 2-pulse CPMG are shown in Fig.~\ref{fig:contour}).
Such an average can be easily performed when the process $\xi_{\text{O}}(t)$ is assumed to be Gaussian (see e.g.~\cite{deSousa_TAP09,Cywinski_PRB08}). A phenomenological derivation of the properties of the $\xi_{\text{O}}$ process was given in \cite{deSousa_PRB03,deSousa_TAP09}. Let us also note that in the situation when the decoherence occurs on times much longer than the typical interaction times between the spins in the bath (e.g.~for dipolar coupling between the central spin and the bath spins), the assumption that $\xi_{\text{O}}(t)$ is an Ornstein-Uhlenbeck process was used to calculate the Rabi oscillations signal \cite{Dobrovitski_PRL09}, and it has been shown to describe very accurately experiments on dynamical decoupling for Nitrogen Vacancy spin in diamond \cite{deLange_Science10}. 

A fully quantum-mechanical solution to the spectral diffusion problem was given during the last 5 years \cite{Witzel_PRB05,Witzel_PRB06,Yao_PRB06,Saikin_PRB07}.  Below we sketch this solution.

Since the diagonal elements of the density matrix of the electron spin are constant in the pure dephasing case, we only need to consider the dynamics of the off-diagonal element $\rho^{S}_{+-}(t)$. The spin is interacting with the bath, so its density matrix is in fact a reduced one, obtained by tracing out the bath degrees of freedom from the total density matrix of the whole system (spin$+$nuclei):
\beq
\rho^{S}_{+-}(t) = \bra{+} \text{Tr}_{\text{B}} \hat{\rho}(t) \ket{-} \,\, ,  \label{eq:rhopm}
\eeq
where $\ket{\pm}$ are the eigenstates of the $\hat{S}^{z}$ operator, and $\text{Tr}_{\text{B}}$ denotes tracing over the nuclear states. We assume a factorizable initial density matrix given by $\hat{\rho}(0) \! = \! \hat{\rho}_{\text{S}}(0) \hat{\rho}_{\text{B}}(0)$. For simplicity we will also assume that $\rho^{S}_{\pm}(0) \! =\! 1$, and we will work with the \emph{decoherence function} $W(t) \! \equiv \! \rho^{S}_{\pm}(t)/\rho^{S}_{\pm}(0)$.
  
The operator $\hat{U}(t)$ describes the evolution  of the whole system under the action of a series of ideal ($\delta$-shaped) $\pi$ pulses applied to the electron spin. For pulses corresponding to rotations by angle $\pi$ about the $\hat{x}$ axis we have
\begin{equation}
\hat{U}(t) = (-i)^{n} \, e^{-i\hat{H}\tau_{n+1}} \hat{\sigma}_{x} e^{-i\hat{H}\tau_{n}} \, ... \, \hat{\sigma}_{x} e^{-i\hat{H}\tau_{1}} 
\end{equation}
with $n$ being the number of applied pulses, $\tau_{i}$ being time delays between the pulses, and the total evolution time $t \! = \! \sum_{i=1}^{n+1} \tau_{i}$ (SE sequence corresponds to $n\!=\!1$ and $\tau_{1} \! = \! \tau_{2} \! = \! t/2$). 

Plugging $\hat{\rho}(t) \! = \! \hat{U}(t)\hat{\rho}(0)\hat{U}^{\dagger}(t)$ into Eq.~(\ref{eq:rhopm}) we arrive at the expression for the FID decoherence function
\beq
W_{\text{FID}}(t) =   \left \langle e^{i\hat{H}_{-}t} e^{-i\hat{H}_{+}t} \right \rangle  \label{eq:WU_FID} 
\eeq
with
\beq
\hat{H}_{\pm} = \hat{H}_{\text{Z}} + \hat{H}_{\text{dip}} \pm \frac{1}{2}\sum_{i}A_{i} \hat{J}^{z}_{i} \,\, ,
\eeq
and $\langle ... \rangle$ denoting the trace over the nuclei.
Similarly, for SE we have
\beq
W_{\text{SE}}(t) =   \left \langle e^{i\hat{H}_{+}\tau}e^{i\hat{H}_{-}\tau} e^{-i\hat{H}_{+}\tau}e^{-i\hat{H}_{-}\tau} \right \rangle  \,\, , \label{eq:WU_SE} 
\eeq
where $\tau\! = \! t/2$. Analogous expressions are obtained for DD sequences with more pulses. 

Equations (\ref{eq:WU_FID}) and (\ref{eq:WU_SE}) can be interpreted in the following way: we have to calculate the evolution of the nuclei under a time-dependent Hamiltonian (in which the sign of the Overhauser term is changing at the times at which the pulses are applied), first forward in time, and then backwards in time, but with a \emph{different} Hamiltonian. The presence of the electron spin is only causing $\hat{H}_{+} \! \neq \! \hat{H}_{-}$. 

By going over to the interaction picture with respect to $\hat{H}_{\text{dip}}$ (slightly complicated by the presence of $\pi$ pulses, see \cite{Cywinski_PRB09}), 
and after introducing the notion (well-known in the theory of nonequilibrium quantum dynamics, see e.g.~\cite{Rammer})  of ordering of operators along a closed time-loop contour $\mathcal{C}$, shown in Fig.~\ref{fig:contour}, the expression for $W(t)$ can be written in a very compact form
\beq
W(t) = \left\langle \mathcal{T}_{\mathcal{C}} \exp \left( -i  \int_{\mathcal{C}} \mathcal{\hat{H}}_{\text{dip}}(\tau_{c}) d\tau_{c} \right)     \right\rangle  \,\, , \label{eq:WC}
\eeq
where $ \mathcal{T}_{\mathcal{C}}$ is the contour-ordering operator, and $\tau_{c} \!= \! (\tau,c)$, with $\tau$ being the time variable and $c \! =\! \pm$ being the contour branch label (see Fig.~\ref{fig:contour}). The operators within the $\mathcal{\hat{H}}_{\text{dip}}(\tau_{c})$ interaction are given by
\beq
\hat{J}^{\pm}_{k}(\tau_{c})  =  \hat{J}^{\pm}_{k} \exp\left [ \pm i \omega_{k}\tau \pm i c \int_{0}^{\tau}f(t;t')\frac{A_{k}}{2} dt' \right ] \,
\label{eq:JpJm}
\eeq
For more details see \cite{Cywinski_PRB09}. The most important point is that in this Equation we have transformed our task into a calculation of an average of a  generalized exponent (note also the similarity to the expression (\ref{eq:classical})). This can be done for any Hamiltonian of the pure dephasing type (see e.g.~\cite{Lutchyn_PRB08} for an application of this formalism to decoherence of a superconducting qubit).
Such a task is very common in many areas of theoretical physics - a calculation of a partition function of a given  system being one example. Out of many methods devised for dealing with averaging generalized exponents, for our purpose the most useful is the \emph{linked cluster theorem} (or the cumulant expansion) \cite{Kubo_JPSJ62,Negele}. According to this theorem $W(t)$ can be written as an exponent of certain bath  averages of products of $\mathcal{\hat{H}}_{\text{dip}}$ operators: 
%$W(t) = \exp \left ( \sum_{c} v_{c}  \right ) $.
\beq
W(t) = \exp \left ( \sum_{c} v_{c}  \right )   \,\, , \label{eq:Wlinked}
\eeq
where the $v_{c}$ terms (with the label $c$ denoting sets of nuclear indices) are such that each of them cannot be written as a product of averages involving subsets of nuclei from the set $c$.

It turns out that in order to reliably calculate the NFID or SE signal \emph{due to the dipolarly-induced dynamics} in III-V QDs it is enough to include only the second order $v_{2}$ terms (i.e.~the terms involving the dynamics of pairs of nuclei) in the linked cluster expansion of Eq.~(\ref{eq:WC}). This result has been obtained with a few methods \cite{Witzel_PRB05,Witzel_PRB06,Yao_PRB06,Saikin_PRB07,Yang_CCE_PRB08}. The inclusion of higher order terms was shown to be necessary only when considering certain dynamical decoupling sequences \cite{Witzel_CDD_PRB07,Lee_PRL08}.

The formula for $v_{2}$ in the SE case is obtained by expanding Eq.~(\ref{eq:WC}) to the second order, using Eq.~(\ref{eq:JpJm}), and performing the trace. The result is
\beq
v^{SE}_{2} = -\sum_{i > j} \frac{32b^{2}_{ij}}{A^{2}_{ij}} \sin^{4}\frac{A_{ij}t}{8} \,\, ,  \label{eq:vSE}
\eeq
where we have summed over the pairs of nuclei of the same isotope (the difference of Zeeman energies very strongly suppresses the dipolar interactions between heteronuclear pairs at high $B$ fields). This result can also be obtained by calculating the spin echo signal for only two nuclei coupled to the central spin, expanding the result to the lowest order in $b_{ij}$, and summing over all the nuclear pairs. 
%One can also include the effect of the dipolar linewidth of the nuclei \cite{Yao_PRB06,Liu_NJP07} by adding the contributions of the other nuclei to the energies of a given pair (due to the $\hat{J}^{z}_{i}\hat{J}^{z}_{k}$ dipolar interactions)  -  TOO MUCH DETAILS???

The above discussion was an attempt to outline the basic structure of the quantum theory of spectral diffusion. The reader interested in technical details  is referred to the original papers \cite{Witzel_PRB05,Witzel_PRB06,Yao_PRB06,Saikin_PRB07} (among which \cite{Yao_PRB06} is probably the most accessible for a newcomer to the subject, and the most detailed discussion of the diagrammatic linked cluster theorem for dipolar interactions is given in \cite{Yang_CCE_PRB08}). Here let us only state the results of calculations for GaAs dots. In these we have a large number $N$ of relevant nuclei, and while each term in the sum in Eq.~(\ref{eq:vSE}) is very small, their sum is large enough to cause decoherence on a timescale of $\sim \! 10 $ $\mu$s (that is why it is enough to consider the pairs only). On this timescale we can use an approximation $\sin A_{ij}t/8 \approx A_{ij}t/8$ in Eq.~(\ref{eq:vSE}). This, together with Eq.~(\ref{eq:Wlinked}), leads to prediction of the $\exp( -(t/T_{\text{SE}})^{4})$ spin echo decay for GaAs. The characteristic time  $T_{\text{SE}} \! =\! 10-100$ $\mu$s depending on the shape and the size of the dot \cite{Witzel_PRB06,Yao_PRB06,Witzel_PRB08}. The NFID decay time was calculated to be of the same order of magnitude \cite{Yao_PRB06}. A significant enhancement of the coherence time was predicted under dynamical decoupling \cite{Yao_PRL07,Witzel_PRL07,Witzel_CDD_PRB07,Lee_PRL08}, with the decay time prolonged from $\approx \! 30$ $\mu$s for SE to $\approx \! 300$ $\mu$s under a 6-pulse CPMG sequence in a GaAs QD \cite{Witzel_PRL07}.

As mentioned in the Introduction, these theories of SE decay due to spectral diffusion were successfully tested in measurements on spins of electrons bound to donors in Si. Very recently, the SE decay was measured in a GaAs-based qubit  for a wide range of magnetic fields \cite{Bluhm_NP11}. The decay at $B \! \gg \! 0.1$ T was seen to be $B$-independent, and to be well-fit by $\exp( -(t/T_{\text{SE}})^{4})$  dependence with $T_{\text{SE}} \! \approx \! 37$ $\mu$s. These observations are in a very good agreement with the theoretical predictions discussed above.

%%%%%%%%%%%%%%%%%%%%%%
%%%% HF AT LOW FIELDS
%%%%%%%%%%%%%%%%%%%%%%
\subsection{6. Hyperfine-induced dynamics in moderate and low magnetic fields}  \label{sec:hf_only}
While the dipolarly-induced dynamics of the nuclei practically does not depend on the magnitude of the magnetic field, the role of the processes involving the $\hat{V}_{\text{ff}}$ part of the hf interaction is growing with decreasing $B$. This is most easily seen by looking at the $\Omega$ dependence of the hf-mediated interaction derived in Section 2. At certain value of $\Omega$ it is expected that the electron spin decoherence will be dominated by purely hf-induced dynamics. 

Initial attempts \cite{Khaetskii_PRB03} to construct the decoherence theory employing the full hf Hamiltonian (and going beyond the QSBA) were successful only in special cases (e.g.~of nearly fully polarized bath), and standard time-dependent perturbation theory was shown to be of very limited use in this problem. Significant progress was made later by using the Generalized Master Equation approach \cite{Coish_PRB04,Coish_PRB10} and the equations of motion method for spin correlation functions \cite{Deng_PRB06,Deng_PRB08}. These approaches were used only in case of NFID, and assuming rather high magnetic fields ($\mathcal{A}/\Omega \! < \! 1$ corresponding to $B \! > \! 3$ T in GaAs).  

The use of an approximate effective Hamiltonian containing the hf-mediated interactions allows for construction of a theory which is structurally similar to the successful theory of spectral diffusion. We simply replace the dipolar interactions by the hf-mediated interaction in Eq.~(\ref{eq:WC}), with the only formal complication being the fact that the hf-mediated interaction from Eq.~(\ref{eq:H2}) is $\hat{S}^{z}$-conditioned (see \cite{Cywinski_PRB09} for details). 

There is one crucial difference with respect to the spectral diffusion theory. There, one only had to retain terms associated with small clusters of nuclei (e.g.~pairs in the case of SE) in the cluster expansion of $W(t)$. This was related to the fact that the dipolar interaction quickly decays with the distance between the nuclei, and at the timescale of the coherence decay the multi-spin nuclear correlations do not have the time to build up. The situation is very different for hf-mediated interactions, which are very long-ranged, and one should include clusters of all possible sizes in the calculation. The assumption that only pairs contribute is valid only at high $B$ fields (when $\mathcal{A}/\Omega \! \ll \! 1$), at which such Pair Correlation Approximation (PCA) calculations were first performed for the hf-mediated interactions \cite{Yao_PRB06,Liu_NJP07,Yao_PRL07}.

Luckily, there is a natural solution to this problem. When all the $N$ spins are comparably coupled to each other, $1/N$ becomes a small parameter which controls the magnitude of contributions of various diagrams in the linked cluster expansion. It is possible then to sum all the diagrams of the leading order in $1/N$, the so-called \emph{ring diagrams}, and to obtain closed solutions for decoherence under any sequence of pulses \cite{Cywinski_PRL09,Cywinski_PRB09}. Again, the formal feature of dealing with an average of a generalized exponent is crucial: the solution is analogous to the calculation by $1/z$ expansion of the partition function of the Ising model with long-range interactions \cite{Brout_PR60} (where $z$ is the number of spins appreciably coupled with each other).

Within this Ring Diagram Theory (RDT) one obtains \cite{Cywinski_PRL09,Cywinski_PRB09}
\beq
W(t) \approx \exp \left( \sum_{k=2}^{\infty} \frac{(-i)^{k}}{k} R_{k}(t) \right)  \,\, , \label{eq:W_R}
\eeq
where $R_{k}$ are the expressions for ring diagram averages. For NFID we have (assuming a single nuclear isotope for simplicity)
\begin{eqnarray}
R_{k} & = & \left (\frac{2}{3}J(J+1)\right)^{k} \int dA_{1} ... \int dA_{k} \rho(A_{1}) ... \rho(A_{k}) \,  \nonumber\\
& & \times \frac{A^{2}_{1} ... A^{2}_{k}}{(2\Omega)^{k}} \frac{\sin A_{12}t}{A_{12}} \frac{\sin A_{23}t}{A_{23}} \, ... \,  \frac{\sin A_{k1}t}{A_{k1}} \,\, ,
\end{eqnarray}
where $\rho(A)$ is the distribution of the hf couplings. At short times $t \! \ll \! N/\mathcal{A}$ we get $R_{k}(t) \approx (\eta t)^{k}$, and the summation of the series in Eq.~(\ref{eq:W_R}) gives us the previously obtained QSBA solution from Eq.~(\ref{eq:SxFID_QSA}).
% SOME MORE DETAILS INTHE SUMMATION? EXPANSIONS OF ATAN AND ln(1+x^2)???
On the other hand, at long times $t \! \gg \! N/\mathcal{A}$ we obtain an exponential decay of NFID signal, $W(t) \! \approx\! \exp(-t/T_{2})$, with $T_{2} \! \sim \! N\Omega^{2}/\mathcal{A}^{3}$ for $\mathcal{A}/\Omega \! \ll \! 1$ (this was also obtained in \cite{Liu_NJP07} and \cite{Coish_PRB08}).
At these high $B$ fields this solution agrees with the one obtained by a very different theory using the full hf Hamiltonian \cite{Coish_PRB10}. Thus, using the effective Hamiltonian and the RDT we can reproduce the QSBA at short times, and also the long-time exponential decay which becomes important at high $B$ fields (the $1/t^{2}$ decay at even longer times predicted in \cite{Deng_PRB06,Deng_PRB08,Coish_PRB10} is however not recovered by the RDT).

The earlier PCA theory of decoherence due to hf-mediated interactions \cite{Yao_PRB06,Liu_NJP07,Liu_AP10} closely corresponds to retaining only the $R_{2}$ term in the sum in Eq.~(\ref{eq:W_R}). This is correct at high $B$ fields ($\Omega \! \gg \! \mathcal{A}$), at which the calculations in these papers were performed. 

For SE decay the RDT gave very characteristic predictions \cite{Cywinski_PRL09,Cywinski_PRB09}. At very high $B$ fields the hf-mediated interactions between nuclei of different species are suppressed due to the differences Zeeman energies $\omega_{\alpha\beta} \! =\! \omega_{\alpha}-\omega_{\beta}$. If one assumes a complete suppression of such inter-species interactions, the result is the absence of any decay of the SE signal \cite{Yao_PRB06}. This can be quickly checked by plugging the Hamiltonian with only intra-species hf-mediated interaction into Eq.~(\ref{eq:WU_SE}). Since such an interaction term commutes with the nuclear Zeeman term, the four exponents cancel each other. In order to obtain the SE decay one has to sum all the ring diagrams which contain only the inter-species interactions. The result is particularly simple in the short time regime ($t \! \ll \! N/\mathcal{A}$):
\beq
W_{\text{SE}}(t) =   \frac{1}{1+R(t) } \,\, ,  \label{eq:W_SE_R2}
\eeq
with the function $R(t)$ given by
\beq
R(t) = \sum_{\alpha\neq\beta} \frac{4\mathcal{A}^{2}_{\alpha}\mathcal{A}^{2}_{\beta}}{N^{2}\Omega^{2}\omega^{2}_{\alpha\beta}} a_{\alpha}a_{\beta}n_{\alpha}n_{\beta}  \sin^{4} \frac{\omega_{\alpha\beta}t}{4}  \,\, ,  \label{eq:R}
\eeq
where $a_{\alpha} \! = \! \frac{2}{3}J_{\alpha}(J_{\alpha}+1)$.

\begin{figure}[t]
\centering
\includegraphics[width=0.99\linewidth]{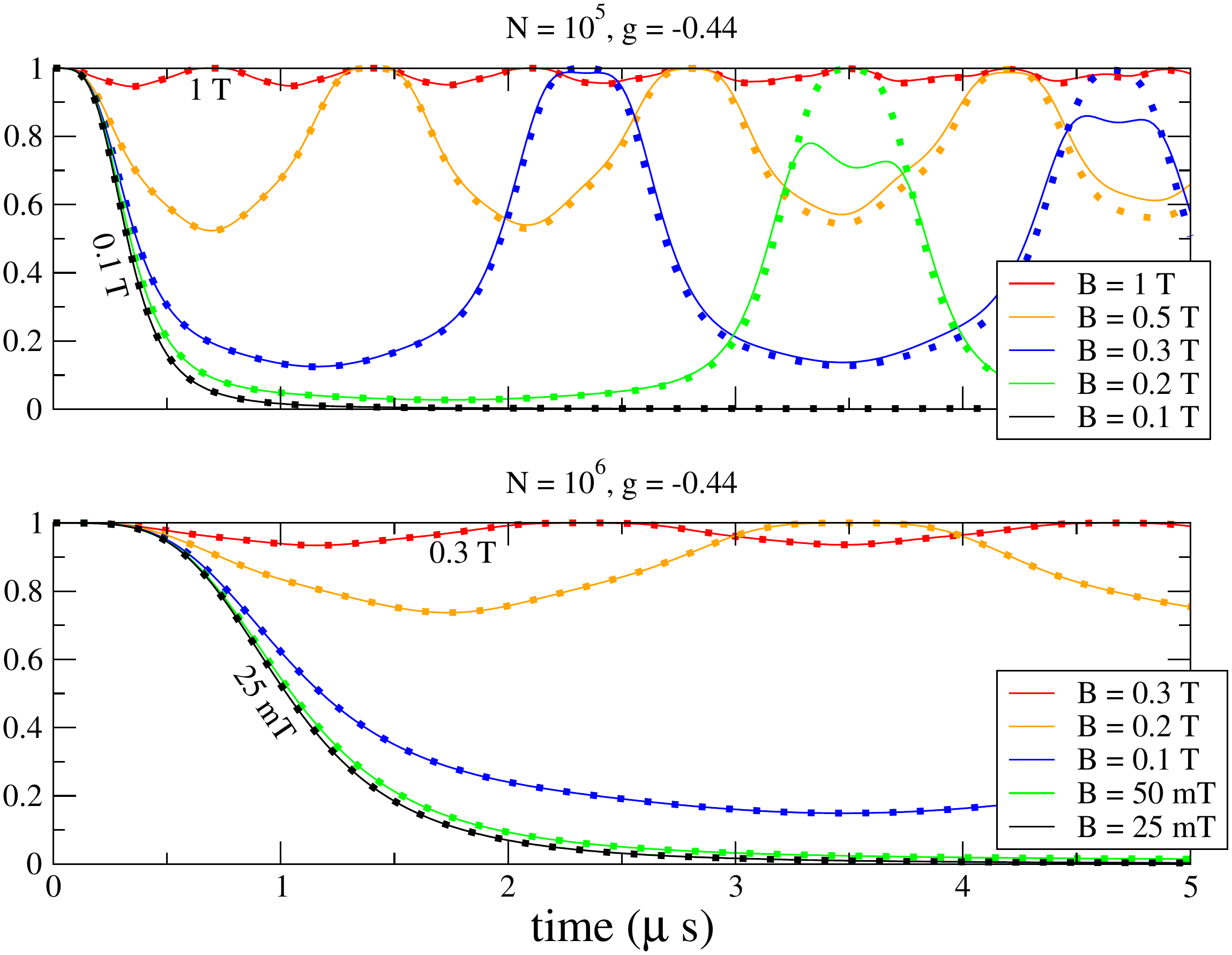}
  \caption{Spin echo coherence function $W_{\text{SE}}$ in GaAs QD with $N\! = \! 10^{5}$ and $10^{6}$ nuclei. The dots are calculated using an approximate analytical expression from Eqs.~(\ref{eq:W_SE_R2}) and (\ref{eq:R}), while lines are the result of the numerical summation of the ring diagrams (see \cite{Cywinski_PRB09} for details). The SE decay due to spectral diffusion (see Section 5) occurs on timescale of  $T_{\text{SE}}^{\text{sd}} \! > \! 10$ $\mu$s for these dots, and the decay in a SE real dot is given by a product of the results from this figure and $\exp[-(t/T_{\text{SE}}^{\text{sd}})^{4}]$ factor. Such a decay was observed in recent experiments \cite{Bluhm_NP11}. The figure is adapted from Ref.~\cite{Cywinski_PRL09}.} \label{fig:SE}
\end{figure}

The SE signals calculated using the RDT are shown in Fig.~\ref{fig:SE} for GaAs dots at different magnetic fields. At high $B$, the hf-mediated interactions are causing the SE signal to oscillate with frequencies related to differences of nuclear Zeeman frequencies. Note that because $\omega_{\alpha\beta}$ are not commensurate, the signal is not strictly periodic, and its revivals  are never complete.
With decreasing $B$ the amplitude of these oscillations grows, and the characteristic period becomes longer. Below the $B_{c}$ field corresponding to the electron Zeeman splitting $\Omega_{c} \! \approx \! \sqrt{r} \mathcal{A}/N$ (where $r$ is the ratio of electronic and nuclear Zeeman energies, $r\approx 10^{3}$), the SE signal goes practically to zero on the timescale of $T_{\text{SE}} \! \approx \! 3\sqrt{r} T^{*}_{2}$, which in GaAs translates to $T_{\text{SE}} \! \approx \! 100 T^{*}_{2}$. The partial revival of the signal at long times is then suppressed by the decay due to spectral diffusion, and this $T_{\text{SE}} $ can be considered a time of low-$B$ irreversible decay of the Spin Echo.

The decay of the SE signal at such a timescale at low $B$ was observed in earlier experiments \cite{Petta_Science05,Koppens_PRL08}. Very recently, the SE was measured in a GaAs singlet-triplet qubit for a wide range of $B$ fields \cite{Bluhm_NP11}. The characteristic oscillations and their evolution with $B$ field was clearly observed there, and a very good fit of the RDT calculation to these results was shown. At $B\! > \! 0.3 $ T the SE decay due to the hf-mediated interactions was suppressed, and the decoherence was occurring due to the spectral diffusion.

Finally, let us touch upon the question of limits of applicability of the RDT. 
Based on the analysis of the higher-order terms in the effective Hamiltonian, in \cite{Cywinski_PRB09} it was conjectured that this theory is reliable for magnetic fields much larger than the rms of the Overhauser field (i.e.~for $\Omega \!  \gg \! \mathcal{A}/\sqrt{N}$, which for a typical GaAs dot translates into $B$ much larger than a few mT), \emph{at least at short times} $t\! \ll \! N/\mathcal{A}$  (corresponding to $t \! \ll \! 10 $ $\mu$s in GaAs). 
This conjecture can be considered experimentally confirmed in the case of SE \cite{Bluhm_NP11}. The accuracy of the RDT for the SE calculation at low $B$ fields was also confirmed by a study in which the RDT results were compared with the exact numerical simulations for a system of $N\! =\! 20$ spins \cite{Cywinski_PRB10}, where it was shown that general predictions of the RDT (e.g. a qualitative difference between hetero- and homo-nuclear baths) hold even at $\Omega \!  \approx \! \mathcal{A}/\sqrt{N}$.

At much higher fields (when $\Omega \! \gg \! \mathcal{A}$) the RDT agrees with other theories \cite{Coish_PRB08,Coish_PRB10} at much longer times. The fact that in $t\! \ll \! N/\mathcal{A}$ limit the RDT corresponds to the QSBA in the case of NFID, together with the success of QSBA in description of Rabi oscillation experiments at these times, supports our statement that low-$B$ and short-time behavior is correctly captured by the RDT.

\subsection{7. Summary and conclusions}
The dephasing of an electron spin by the nuclear bath is a nontrivial theoretical problem because of the strong (relative to the Zeeman energies at low magnetic fields) hyperfine coupling between the electron and the nuclei, and because of the slow dynamics of the nuclei. In fact, many experiments  (Free Induction Decay for example) can be explained by a theory assuming that the nuclei are static. After a preparation of a special ``narrowed'' state of nuclei (in which the inhomogeneous broadening is decreased), or in a Spin Echo experiment, the coherence decay occurs at longer timescales, and the dynamics of the nuclei has to be considered. At high magnetic fields (larger than a few hundreds of mT in typical GaAs dots) the nuclei are fluctuating due to the dipolar interactions between them. The quantum theory of decoherence due to this process (the so-called spectral diffusion) successfully explains the results of Spin Echo experiments at high fields \cite{Witzel_PRB05,Witzel_PRB06,Yao_PRB06,Saikin_PRB07}. At lower fields (between $\sim \! 10$ and a few hundreds of mT in GaAs) the hyperfine-mediated interactions among the nuclei are the dominant source of the nuclear dynamics causing the electron spin dephasing. The theoretical approach to this problem had given predictions for low-field Spin Echo decay \cite{Cywinski_PRL09,Cywinski_PRB09}, which have been recently confirmed experimentally \cite{Bluhm_NP11}. 
 
\subsection{Acknowledgements}
The author would like to thank his collaborators: W.M.~Witzel, V.V.~Dobrovitski, X. Hu, and S.~Das Sarma. Enlightening discussions with W.A.~Coish are also greatly appreciated. The financial support from the Homing programme of the Foundation for Polish Science supported by the EEA Financial Mechanism is gratefully acknowledged. Some of the work reviewed here was done at the University of Maryland where it was supported by the LPS-NSA-CMTC grant

\end{document}